\documentclass[pra,twocolumn,showpacs]{revtex4}
\usepackage{amsmath}
\usepackage{amssymb}
\usepackage{epsfig}

\begin{document}

\title{\bf Long-lived quantum vortex knots}
\author{V.~P. Ruban}
\email{ruban@itp.ac.ru}
\affiliation{L.D. Landau Institute for Theoretical Physics RAS, Moscow, Russia} 
\date{\today}

\begin{abstract}
Dynamics of simplest quantum vortex knots of torus type in a superfluid at zero temperature 
has been simulated within a regularized Biot-Savart law (the torus radii $R_0$ and $r_0$
for initial vortex configuration were large in comparison with a vortex core width $\xi$). 
Computations of evolution times of knots until their significant deformation were 
carried out with a small step on parameter $B_0=r_0/R_0$ for different values of parameter 
$\Lambda=\log(R_0/\xi)$. It has been found that at $\Lambda\gtrsim 3$, bands of quasi-stability 
appear in a region of  $B_0\lesssim 0.2$, which correspond to long knot lifetimes and to very large 
traveling distances --- up to several hundreds of $R_0$. This result is new and quite unexpected, 
because previously it was believed that maximal lifetime of torus knots until reconnection does 
not exceed several typical periods. The opening of quasi-stable 'windows' at increasing $\Lambda$ 
is due to narrowing main parametric resonances of the dynamical system on parameter $B_0$.
\end{abstract}
\pacs{47.37.+q, 67.25.dk, 03.75.Kk, 67.85.De}
\maketitle

{\bf Introduction}.
In the bulk of a superfluid, besides well-known and experimentally observed quantum vortex rings,
theoretically there can exist (developing in time) also solitary topologically non-trivial excitations
as vortex knots (see \cite{RSB999,MABR2010,POB2012,KI2013,POB2014,LMB016,KKI2016}, and references therein).
The simplest are torus knots ${\cal T}_{p,q}$, where  $p$ and $q$ are co-prime integers.
The corresponding tor at an initial time moment is determined by two sizes, the toroidal (large) radius 
$R_0$ and the poloidal (small) radius $r_0$, both of them being supposed large in comparison with a width 
of quantum vortex core $\xi$. It is believed on the basis of previously obtained numerical results
(see the above cited references) that torus knots are unstable and they reconnect during just few typical 
times, traveling a distance in several $R_0$ (the lifetime is somewhat longer for smaller ratios 
$B_0=r_0/R_0$). Due to this property, such knots are often taken as initial conditions to study 
reconnection of vortex lines. The mentioned results were obtained for not too large ratios 
$R_0/\xi\lesssim 20$, and with a very coarse step (about 0.1) on parameter $B_0$. 

\begin{figure}
\begin{center}
\epsfig{file=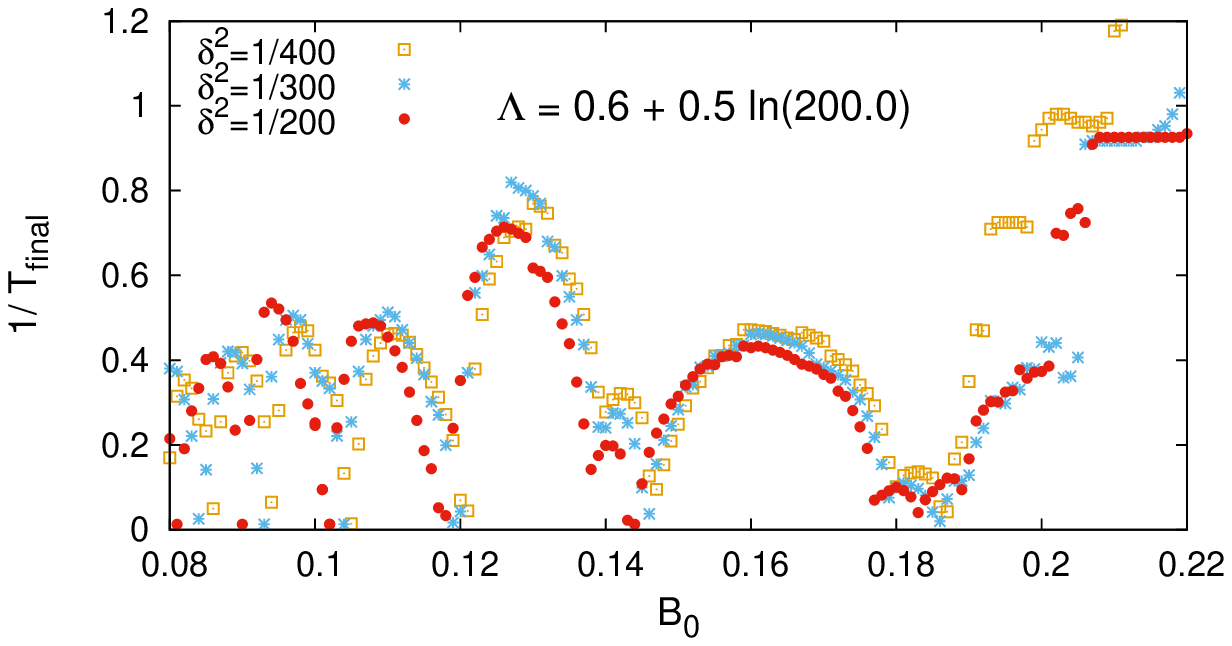, width=80mm}\\
\epsfig{file=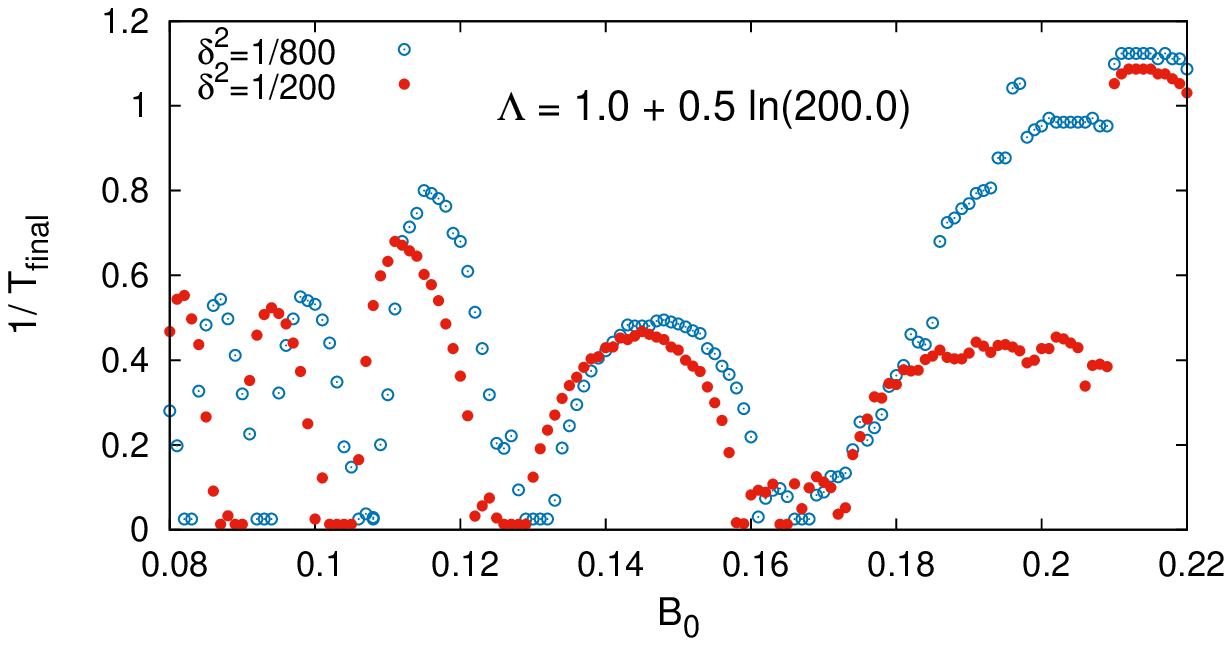, width=80mm}\\
\epsfig{file=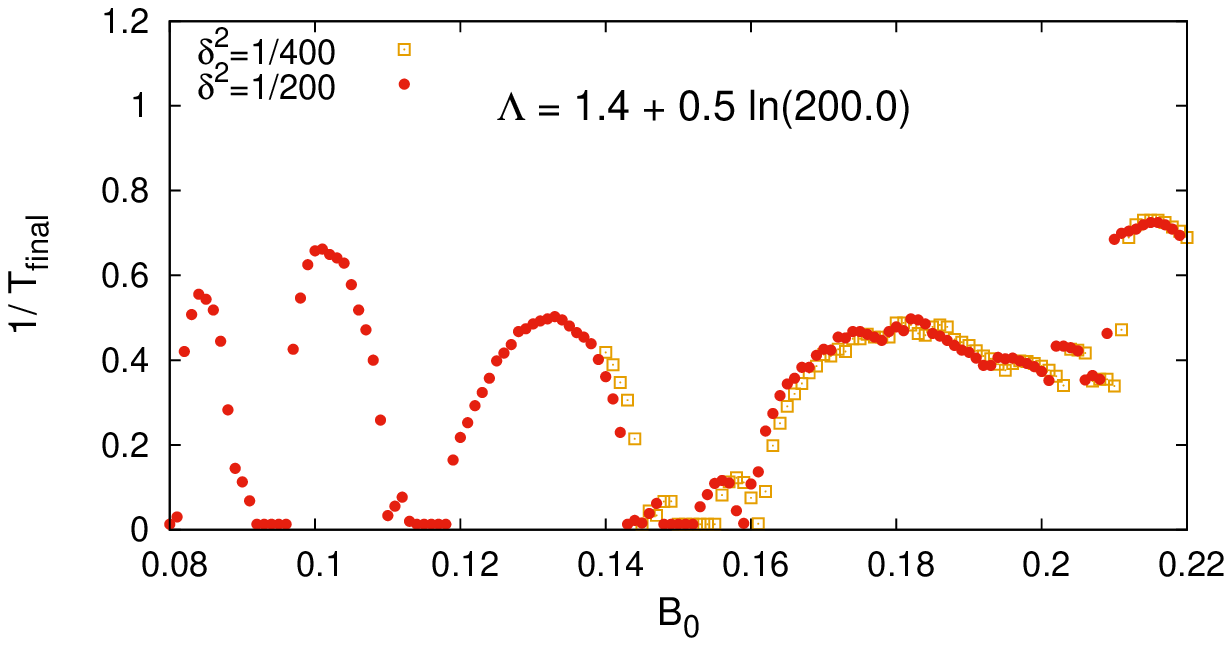, width=80mm}\\
\epsfig{file=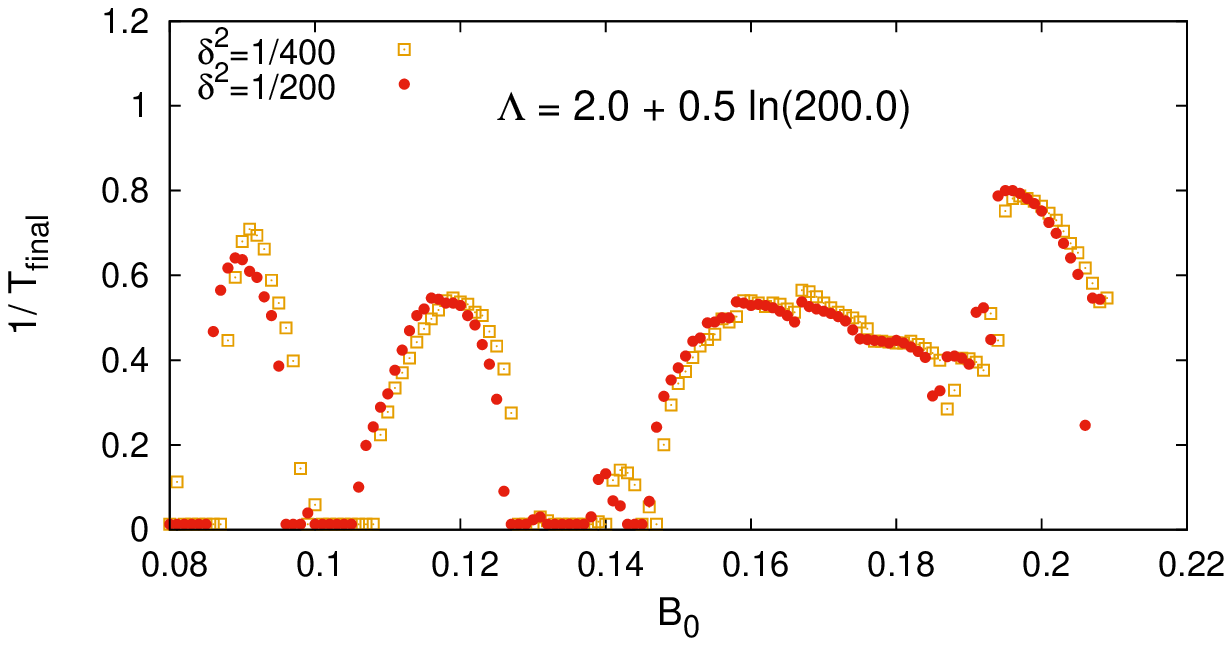, width=80mm}
\end{center}
\caption{The inverse lifetime of trefoil ${\cal T}_{2,3}$ vortex knot.}
\label{trefoil} 
\end{figure}

In this work by the example of trefoil knot  ${\cal T}_{2,3}$ it is shown that actually the situation
is much more complicated and interesting. Namely, at fixed values of parameter $\Lambda=\log(R_0/\xi)$,
the dependence of knot lifetime on parameter $B_0$  turns out to be drastically non-monotonic on sufficiently
small $B_0\lesssim 0.2$. Moreover, at $\Lambda\gtrsim 3$ quasi-stability bands appear, where vortex knot
remains nearly unchanged for many dozens and even hundreds of typical times (see Fig.\ref{trefoil}).
The revealing of quasi-stable domains in the parametric space of vortex knots seems to be an important
theoretical result. Let us below describe the way how it was obtained.

\begin{figure}
\begin{center}
\epsfig{file=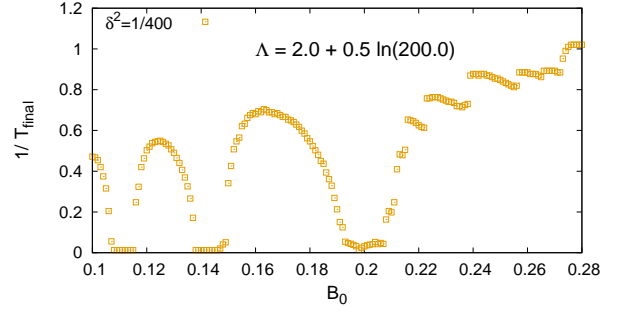, width=80mm}
\end{center}
\caption{The inverse lifetime of vortex knot ${\cal T}_{3,2}$.}
\label{T32} 
\end{figure}

{\bf Model}.
In the low-temperature limit, long-wave dynamics of a quantum vortex filament in a superfluid can be
approximated with a good accuracy by a regularized Biot-Savart law (see, e. g., \cite{S1985,TAN2000,BB2011}
and many references therein), together with a local induction contribution,
\begin{equation}
{\bf X}_t(\beta,t)=\frac{\Gamma}{4\pi}\oint \frac{{\bf X}_1'\times({\bf X}-{\bf X}_1)}
{\mbox{reg}_\xi|{\bf X}-{\bf X}_1|^3}d\beta_1+ \frac{\Gamma\Lambda_0}{4\pi}\varkappa{\bf b},
\label{BS_reg}
\end{equation}
where $\beta$ is an arbitrary longitudinal parameter along the curve, $t$ is the time,
$\Gamma=2\pi\hbar/m$ is the velocity circulation quantum ($m$ being the atomic mass), 
${\bf X}_1={\bf X}(\beta_1,t)$, ${\bf X}_1'=\partial {\bf X}(\beta_1,t)/\partial\beta_1$, 
$\Lambda_0$ is a dimensionless positive parameter characterizing potential energy of vortex core,
$\varkappa$ is a local curvature of the filament, ${\bf b}$ is a local unit binormal vector. 
Dynamics of filament is almost insensitive to a manner of regularization of the logarithmically 
diverging integral, provided the parameter $\Lambda_0$ is defined consistently. Often the choice is made,
\begin{equation}
\mbox{reg}_\xi|{\bf X}-{\bf X}_1|^3=\sqrt{(|{\bf X}_1-{\bf X}_2|^2+\xi^2)^3}.
\label{reg}
\end{equation}

As an additional argument for applicability of Eq.(\ref{BS_reg}), we can say about a number of
quasi-microscopic models of superfluid hydrodynamics in the form of a modified Gross-Pitaevskii equation
\cite{BR1999,BBP2014,AS2008}
\begin{equation}
i\hbar\Psi_t={\delta H}/{\delta\Psi^*},
\label{Psi_variat}
\end{equation}
with different Hamiltonians of the type
\begin{equation}
H\{\Psi,\Psi^*\}=\int \frac{\hbar^2}{2m}|\nabla\Psi|^2d^3{\bf r}+ W\{|\Psi|^2\},
\end{equation}
where the potential energy functional $W\{|\Psi|^2\}$ may be non-local as well as local.
In those models the complex order parameter $\Psi=\sqrt{\rho_s/m}\exp(i\Phi)$ contains
superfluid density $\rho_s$ and superfluid velocity ${\bf v}_s=(\hbar/m)\nabla\Phi$.
It follows from some results of recent work \cite{BN2015} that for such dynamical systems 
the corresponding equation of motion for a vortex filament possesses the following non-canonical 
Hamiltonian structure in the long-scale limit,
\begin{equation}
\Gamma[{\bf X}'\times{\bf X}_t]\rho_0\approx \delta{\cal H}/\delta{\bf X(\beta)},
\label{variat_X}
\end{equation}
where $\rho_0$ is an equilibrium density of superfluid far away from the vortex.
The dynamics (\ref{BS_reg}) follows from this equation with a definite form of the longitudinal 
component of vector ${\bf X}_t$, which component remains undetermined within Eq.(\ref{variat_X})
in correspondence with the freedom of longitudinal parameterization $\beta$. It is implied here that
Hamiltonian ${\cal H}$ of the filament is equal to sum of a vortex-created kinetic energy
${\cal K}=\int(\rho_s{\bf v}_s^2/2) d^3{\bf r}$ and an effective potential energy ${\cal U}$ 
(with taking into account the quantum pressure), the last term being concentrated in the vortex core.
The functional ${\cal K}$ is non-local, and with the choice of regularization as
(\ref{reg}) it takes form
\begin{equation}
{\cal K}\{{\bf X}(\beta)\}=\frac{\rho_0\Gamma^2}{8\pi}\oint\oint 
\frac{({\bf X}_1'\cdot{\bf X}_2') d\beta_1 d\beta_2}{\sqrt{|{\bf X}_1-{\bf X}_2|^2+\xi^2}}.
\end{equation}
The functional ${\cal U}$ is proportional to the length of vortex filament,
\begin{equation}
{\cal U}\{{\bf X}(\beta)\}=\frac{\rho_0\Gamma^2\Lambda_0}{4\pi}\oint|{\bf X}'|d\beta,
\end{equation}
with parameter $\Lambda_0$ of order unity, exact value of which depends on explicit form of functional 
$W$. This parameter gives stiffness to short-wave modes.

It is important that long-scale dynamics of this system is almost insensitive to the parameter change
in the regularized Hamiltonian $\xi\to\delta$, $\Lambda_0\to\Lambda_0+\log(\delta/\xi)$, with
$\delta$ being an arbitrary quantity of order $\xi$, provided a filament configuration is far from
self-intersections. In particular, parameter $\xi$ can be re-defined in such a way that new $\Lambda_0=0$.
This will be supposed below. The above change does not modify the full local induction parameter  
\begin{equation}
\Lambda=\log(R_0/\xi)=\log(R_0/\delta) +\tilde\Lambda_0,
\end{equation}
where $\tilde\Lambda_0=\log(\delta/\xi)$.  
The indicated property of the Hamiltonian allows one to carry out numerical simulations with a shorter 
array of discrete points than it was required by small values of $\xi/R_0$, provided $\delta$ greater than 
several $\xi$ is taken. In our numerical experiments, the main parameters were  $\Lambda$ and $\delta$.

Dimensionless variables were used, so that $\Gamma=2\pi$, $R_0=1$. Thus, dynamical system equivalent 
to equation 
\begin{equation}
{\bf X}_t(\beta,t)=\frac{1}{2}\oint \frac{{\bf X}_1'\times({\bf X}-{\bf X}_1)d\beta_1}
{\sqrt{(|{\bf X}-{\bf X}_1|^2+\delta^2)^3}}+ \frac{\tilde\Lambda_0}{2}\varkappa{\bf b}
\label{BS_reg_delta}
\end{equation}
was simulated, with initial data in the form of slightly perturbed trefoil knot with $R_0=1$, $r_0=B_0$.
By the way, that quasi-random small perturbation was in a sense additional, because an ideal torus knot
itself is not exactly a stationary configuration of our dynamical system. In other words, a stationary 
rotating and traveling knot of unknown shape would lie on a deformed torus --- the more deformed one, 
the larger parameter $B_0$ is.

We used a pseudo-spectral scheme on variable $\beta$, and a Runge-Kutta 4th-order scheme for 
time-integration. Advancing in time was terminated when deformation of knot became sufficiently strong.
At that, for each set of parameters the attained time $T_{\rm final}$ was recorded. As our computational 
practice has shown, exact form of the run termination criterion  is not very important for finding
quasi-stationary knot configurations, so we do not describe here all the details.

\begin{figure}
\begin{center}
\epsfig{file=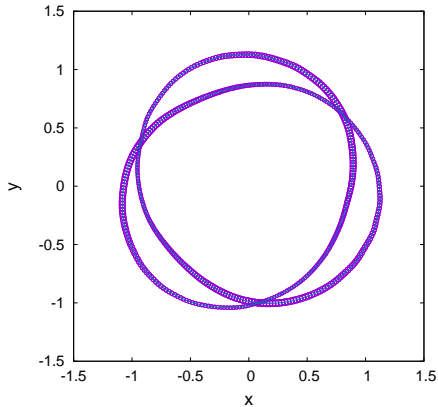, width=60mm}
\end{center}
\caption{Projection of vortex knot ${\cal T}_{2,3}$ with parameters 
$\Lambda=1.4+0.5\log(200)\approx 4.05$, $\delta=1/20$, $B_0=0.154$ on the plane $(x,y)$ at $t=320$.
The line is thicker where local coordinate $z$ is larger.}
\label{T23t320} 
\end{figure}
\begin{figure}
\begin{center}
\epsfig{file=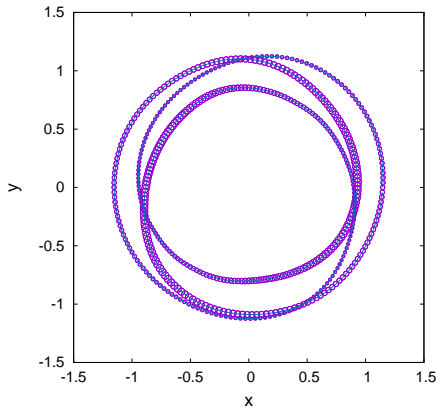, width=60mm}
\end{center}
\caption{Projection of vortex knot ${\cal T}_{3,2}$ with parameters 
$\Lambda=2.0+0.5\log(200)\approx 4.65$, $\delta=1/20$, $B_0=0.198$ on the plane $(x,y)$ at $t=70$.}
\label{T32t70} 
\end{figure}

{\bf Results}.
Numerically found dependencies of the trefoil knot lifetime upon parameter $B_0$ are shown 
in Fig.\ref{trefoil}a-d. It is seen that at fixed $\Lambda$ the difference between results 
corresponding to different $\delta$ is rather small, in accordance with the above remark.
But distinctions are still present, because due to nonlinear interactions, on final times
short-wave filament excitations develop, which feel the difference between Hamiltonians
with different $\delta$. Therefore, maximally accurate modeling needs the original (not re-defined)
parameters  $\xi$ and $\Lambda_0$. This remark however does not cancel the main conclusion about
presence of quasi-stable domains in the parameter plane $(B_0,\Lambda)$.

On the whole, with increase of $\Lambda$ many fragments of the inverse lifetime graphs distinctly 
resemble segments of ellipses with centers on the horizontal axis $B_0$. Some of ellipses intersect, 
but some are separated by intervals where values of $1/T_{\rm final}$ are very small.
Qualitatively similar results take place also for the knot  ${\cal T}_{3,2}$ (see Fig.\ref{T32}).
In fact, in the figures just a lower estimate for the lifetime is presented, because in the windows of
quasi-stability the computations were terminated at a fixed, though quite large $T_{\rm max}$ which moment
was still long before strong knot deformation. It is therefore unclear at the moment if  
the instability increment goes exactly to zero there or it remains finite, though extremely small.
In additional simulations, in some cases the trefoil lifetime exceeded 300, while the distance 
traveled by knot was larger than 1200 (Fig.\ref{T23t320}). The Hamiltonian was conserved in that 
case at least up to 7 decimal digits, which fact manifests good quality of the simulation. 
An example of knot ${\cal T}_{3,2}$ with a long but certainly finite lifetime is shown in Fig.\ref{T32t70}.

If we roughly accept that inverse lifetime found is proportional to the main instability increment,
then such a picture is evidence of parametric resonances present in the system, and they become narrower
as $\Lambda$ increases, while their centers are shifted toward smaller $B_0$. 

{\bf Conclusions}.
Thus, in this work very long-lived quantum vortex knots in a superfluid at zero temperature 
are theoretically predicted in definite domains of geometric parameters. This result essentially 
enriches our knowledge about dynamics of vortex filaments.
An interesting problem for future research is to investigate in detail analytically
the structure of parametric resonances of torus knots as Hamiltonian systems.

\end{document}